\documentclass[a4paper]{article}
\usepackage[latin1]{inputenc}
\usepackage[T1]{fontenc}
\usepackage[english]{babel}
\usepackage{amsfonts}
\usepackage{amssymb}
\usepackage{amsmath}
\usepackage{amsthm}
\usepackage{graphicx}
\usepackage{color}

\newcommand{\meanv}[1]{\left\langle#1\right\rangle}

\def\be{\begin{equation}}
\def\ee{\end{equation}}
\def\bc{\begin{center}}
\def\ec{\end{center}}

\def\s{\sigma}

\def\mb{\mathbb}
\def\p{\partial}

\begin{document}

\title{Mean field spin glasses treated with PDE techniques}


\author{Adriano Barra\footnote{Dipartimento di Fisica, Sapienza Universit\`a di Roma, P.le Aldo Moro 5, Rome, Italy.},
Gino Del Ferraro\footnote{Department of Computational Biology, KTH Royal Institute of Technology, SE-100 44, Stockholm, Sweden.}, \ Daniele Tantari \footnote{Dipartimento di Matematica, Sapienza Universit\`a di Roma, P.le Aldo Moro 5, Rome, Italy.}}
\date{January 2012}

\maketitle

\begin{abstract}
Following an original idea of F. Guerra, in this notes we analyze the Sherrington-Kirkpatrick model from different perspectives, all sharing the underlying approach which consists in linking the resolution of the statistical mechanics of the model (e.g. solving for the free energy) to  well-known partial differential equation (PDE) problems (in suitable spaces). The plan is then to solve the related PDE using techniques involved in their native field and lastly bringing back the solution in the proper statistical mechanics framework.
\newline
Within this strand, after a streamlined test-case on the Curie-Weiss model to highlight the methods more than the physics behind, we solve the SK both at the replica symmetric and at the $1$-RSB level, obtaining the correct expression for the free energy via an analogy to a Fourier equation and for the self-consistencies with an analogy to a Burger equation, whose shock wave develops exactly at critical noise level (triggering the phase transition).
\newline
Our approach, beyond acting as a new alternative method (with respect to the standard routes) for tackling the complexity of spin glasses, links symmetries in PDE theory  with  constraints in statistical mechanics and, as a novel result from the theoretical physics perspective,  we obtain a  new class of polynomial identities (namely of Aizenman-Contucci type, but merged within the Guerra's broken replica measures), whose interest lies in understanding, via the recent Panchenko breakthroughs, how to force the overlap organization to the ultrametric tree predicted by Parisi.
\end{abstract}

\section{Introduction}

More than thirty years elapsed since Parisi gave his solution for the mean-field spin glass \cite{MPV}, namely the Sherrington-Kirkpatrick model (SK). During this period, with continuous joined effort from theoretical, numerical and mathematical physics, the community of researchers involved in the field transformed the SK model into the harmonic oscillator for complex systems and features of glassy phenomenology as multiple time sectors \cite{sectors}, FTD-violations \cite{FDT}, chaos in temperature \cite{rizzo}, weak/strong ergodicity breaking \cite{trap,derrida} and last but not least ultrametricity \cite{MPV} are nowadays considerable building blocks of the scaffold of complexity, whose applications are spreading over disparate disciplines, ranging from economy \cite{bouchaud} and biology \cite{stein} to computer science \cite{mezard} and artificial intelligence \cite{peter2}.
\newline
However, from a rigorous perspective (i.e. avoiding the zero replica limit within the replica trick framework \cite{dotsenko}), only recently, via a series of impressive results achieved by Guerra \cite{broken} and Talagrand \cite{t4} and/or within the A.S.S. scheme \cite{ass,ass2,peter}, a crystal-clear formal understanding of the Parisi expression for the free energy has been achieved, and even more recently \cite{panchenko1,panchenko2,panchenko3,panchenko4,panchenko5}, Panchenko has been able to link Parisi ultrametricity with polynomial constraints as Ghirlanda-Guerra \cite{gg} and Aizenman-Contucci \cite{ac} identities, thus conferring a central importance to the latter.
\newline
Hence, in the actual state of the art, from one side (and mainly for applications) there is a continuous need for new mathematical techniques able to tackle a glassy problem from multiple perspectives and, from another (more conceptual)  side,  a particular focus on the identities (as those we have understood that play a crucial role in the ultrametric organization of the states in the low-noise limit of the SK model) raised.
\newline
Indeed, despite Panchenko breakthroughs in exploiting their deep links with ultrametricity are very recent, polynomial identities have a long history in spin glass theory and, since the beginning, the hierarchical flavor they contain were manifest (as the title of \cite{gg} does not hide): Their early developments are by Ghirlanda and Guerra (GG) \cite{gg} and by Aizenman and Contucci (AC) \cite{ac,guerra2} at the end of the Nineties. Following their seminal approaches, the former based on checking the stability of the thermodynamical states by adding all possible $p$-spin terms \cite{derrida,grem} to the SK Hamiltonian and then sending their strength to zero (self-averaging of the quenched internal energy), the latter obtained trough a property of robustness of the quenched Gibbs measure with respect to small stochastic perturbation \cite{ac,pierluz,contucci} (stochastic stability), identities for the SK model have by now been obtained with a number of different techniques, e.g.\ via smooth cavity field expansion \cite{barra1}, linear response stability \cite{claudio1}, random overlap structures \cite{peter} or even as Noether invariants \cite{BG1}.
\newline
Beyond the full mean-field panorama, it has been possible to show to validity of these polynomial identities even in short-range or finite-dimensional models \cite{claudio2,claudio3,franz,T}, however, already restricting in considering  the SK-model only, the whole identity repertoire is not yet complete and novel techniques to obtain other restrictions, with the aim of finding a set of constraints on the overlap probability distribution strong enough to enforce the replica symmetry breaking scheme, are still a central focus in spin glass theory.
\newline
As a consequence, by merging the interest in always welcome alternative mathematical approaches to the glassy phenomenology with the renewed importance of obtaining novel polynomial identities as they are seen nowadays, we decided to continue the investigation of a way initially paved by F.Guerra in solving the thermodynamics of the SK-model via well-known PDE techniques, as he first exploited this strand via a bridge connecting the  SK free energy with the Hamilton-Jacobi theory \cite{sumrule} and then it has been continued in various extensions as for instance in \cite{BG1}\cite{BarraCW}\cite{BDBG}\cite{shannon}.
\newline
The plan for the present paper is as follows: For the sake of clearness, in the next section (Sec. Two), we outline the techniques we develop by using the Curie-Weiss model as a toy example. At first, we show that the magnetization satisfies a Burger equation whose shock-wave develops exactly at the noise level that triggers the phase transition in the SM language, then we introduce its Cole-Hopf transform that maps the evolution of the magnetization versus the noise level into a diffusive problem described by a Fourier PDE, which is finally solved in a standard way via Green propagator and the Convolution Theorem in the impulse space. As a result of this procedure self-averaging of the order parameter is obtained as a bypass product. Then, in Section Three, we introduce the Sherrington-Kirkpatrick model that we solve immediately at the replica symmetric (RS) level with the same techniques, hence we obtain a Burger equation for the overlap and then we find the shock wave that spontaneously develops at the phase transition where annealing breaks down. Via Cole-Hopf transform we map the latter into a diffusive problem, that we solve obtaining, beyond the replica symmetric expression for the free energy, self-averaging of the order parameter (which is in obvious agreement with the request of a RS-solution) and linear constraints of the Aizenman-Contucci type.
\newline
One step forward, we merge this approach with the classical Guerra's broken replica construction \cite{broken} to go beyond the RS scheme and we show that there is a one-to-one correspondence among the steps of replica symmetry breaking in SM and the spatial dimensions in the equivalent diffusive problem (exactly as happens in the Hamilton-Jacobi framework \cite{BDBG}) that we then solve in all details at the first step of RSB for highlighting the power of the method.
\newline
Remarkably, as a bypass product of this approach, we obtain new identities which constraint overlap fluctuations on different Guerra measures (roughly speaking constraints between different valleys belonging to various steps of RSB). We remark that this kind of identities is completely new and carry interesting physics inside, whose discussion, coupled to general outlooks, constitutes the last section, that closes the present paper.

\section{Testing the machinery: The Curie-Weiss model}

Once introduced $N$ Ising spins $\sigma_i = \pm 1$, with $i = 1, ..., N$, the Hamiltonian of the Curie-Weiss model can be written as
\be
H_N(\sigma) = - \frac{1}{N}\sum_{i<j}\sigma_i \sigma_j \sim -\frac{N}{2} m^2_N,
\ee
where in the last passage (which becomes exact in the thermodynamic limit $N \to \infty$) we introduced the magnetization $m_N = N^{-1}\sum_i^N \sigma_i$, namely the order parameter of the theory.
\newline
We introduce also the Boltzmann averages as, using the magnetization as a trial function
$$
\langle m_N \rangle = \frac{\sum_{\sigma} \frac{1}{N}\sum_i^N \sigma_i \exp\left(-\beta H_N(\sigma)\right)}{\sum_{\sigma} \exp\left(-\beta H_N(\sigma)\right)},
$$
where the denominator is called "partition function" $Z_N(\beta)$ and of course $\lim_{N \to \infty}\langle m_N \rangle = \langle m \rangle$.
\newline
In order to investigate the thermodynamics of the model, we are interested in the mathematical pressure $\alpha(\beta)$ or equivalently in the free energy $f(\beta)$ defined as
\be
\alpha(\beta)= - \beta f(\beta) = \lim_{N \to \infty} \frac{1}{N}\ln Z_N(\beta),
\ee
and we want to achieve an explicit expression for this quantity without following the standard routes of statistical mechanics.
\newline
The approach we want to use is to ''enlarge" the space of the parameters (hence $\beta$), and investigate which PDE are obeyed by the model in such a space, so to import the technology for their resolution from classical mechanics. In order to exploit our idea, let us introduce the following action $S(t,x)$ as
\be
S_N(t,x)=-\frac{1}{N}\ln \sum_{\sigma}\exp\left( \frac{t}{2N}\sum_{ij}\sigma_i \sigma_j + x \sum_i \sigma_i \right),
\ee
where the variables $t,x$ can be thought of as fictitious time and space, and such that $\lim_{N \to \infty}S_N(t,x)=S(t,x)$ and of course $S(t,x) = \beta f(\beta) = -\alpha(\beta)$ whenever evaluated at $t=\beta$ and $x=0$.
\newline
In order to highlight our approach we need to work out the derivatives of $S(t,x)$ which read as
\begin{eqnarray}
\frac{\partial S_N(t,x)}{\partial t} &=& -\frac{1}{2}\langle m^2_N \rangle,\\
\frac{\partial S_N(t,x)}{\partial x} &=&  -\langle m_N \rangle,\\
\frac{1}{2N} \frac{\partial^2 S_N(t,x)}{\partial x^2} &=&  \frac12 \left(\langle m_N^2 \rangle - \langle m_N \rangle^2\right).
\end{eqnarray}


Following the Guerra prescription \cite{sumrule,BDBG} and noticing the form of the derivatives $(4-6)$, it is possible to build an Hamilton-Jacobi equation for $S_N(t,x)$ as
\be\label{CWHJ}
\partial_t S_N(t,x) + \frac12 \left(\partial_x S_N(t,x)\right)^2 + V(t,x) = 0,
\ee
where we remark that $V(t,x) = -(1/2N)\partial^2_{xx}S_N(t,x)= (1/2) (\langle m_N^2 \rangle - \langle m_N \rangle^2)$: In the spin-glass counterpart this will no longer be the case as polynomial constraints will be present (that here simply reduce to e.g. $\langle m^4 \rangle = \langle m^2 \rangle^2$, hence are {\em all} already accounted by the self-averaging of the magnetization).
\newline
Deriving eq.(\ref{CWHJ}) w.r.t. $x$, and calling $u_N(t,x)=\partial_x S_N(t,x)= -\langle m_N \rangle$ we get the following Burger equation \cite{erik3} for the velocity (i.e. the magnetization in the statistical mechanics framework apart the minus sign)
\be\label{CWB}
\partial_t u_N(t,x) + u_N(t,x)\partial_x u(t,x) - \frac{1}{2N}\partial^2_{xx}u_N(t,x)=0,
\ee
and let us point out that such equation becomes naturally inviscid in the thermodynamic limit as $S(t,x)$ admits the thermodynamic limit thanks to the Guerra-Toninelli scheme \cite{limterm}.
\newline
If we now perform the following Cole-Hopf transform 
\be\label{CWcole}
\Psi_N(t,x) = \exp\left( -N \int dx \, u_N(t,x) \right),
\ee
it is immediate to check that $\Psi_N(t,x)$ satisfies the following diffusion equation
\be\label{CWF}
\frac{\partial \Psi_N(t,x)}{\partial t} - \frac{1}{2N}\frac{\partial \Psi_N(t,x)}{\partial x^2}=0,
\ee
which we now solve, in the Fourier space, trough Green propagator and the Convolution Theorem.
In the Fourier space we deal with $\hat{\Psi}_N(t,k)$ defined as
\be
\hat{\Psi}_N(t,k) = \int dk e^{-i k x}\Psi_N(t,x),
\ee
and equation (\ref{CWF}) in the impulse space reads off as
\be
\partial_t \hat{\Psi}_N(t,k) + \frac{k^2}{2N}\hat{\Psi}_N(t,k)=0,
\ee
whose solution is
\be
\hat{\Psi}_N(t,k) = \hat{\Psi}_N(0,k) \exp\left( -\frac{k^2}{2N}t \right),
\ee
and translates in the original space as
\be
\Psi_N(t,x) = \int dy G_t(x-y) \Psi_0(y),
\ee
where the Green propagator is given by
\be
G_t(x-y) = \sqrt{\frac{N}{2 \pi t}}\exp\left( -N (x-y)^2/(2t) \right).
\ee
Overall we get
\be\label{sella}
S(t,x) = -\frac 1 N \ln \sqrt{\frac{N}{2 \pi t}} \int dy\  e^{-N\left( (x-y)^2/(2t)-\ln 2-\ln\cosh(y)\right)},
\ee
where we used $S_N(0,x)=-\log 2 - \ln\cosh(x)$ and the definition of the Cole-Hopf transform (\ref{CWcole}).
\newline
As the exponent in equation (\ref{sella}) is proportional to the volume, for large $N$, we can apply now the saddle point argument to get
\be
\alpha (t,x)= \sup_y\left\{ -\frac{\left( x-y \right)^2}{2t}  + \ln 2 + \ln\cosh(y)\right\} = -\frac{(x-\hat{y})^2}{2t} + \ln 2 + \ln\cosh(\hat{y}),
\ee
with $\hat{y}$ maximizer. Note that the previous equation can also be written as
\be
\alpha(t,x) + \frac{x^2}{2t} = \sup_y\left\{ -\frac{y^2}{2t} + \ln 2 + \ln \cosh (y) + \frac{x y}{t}\right\}=
\sup_y\left\{ \Phi_0(y) + \frac{x t}{y}\right\},
\ee
hence we obtained the solution also as a Legendre transform of the initial condition\footnote{Strictly speaking $\Phi_0(y)= -y^2/2t + \ln 2 + \ln\cosh(y)$, hence is the initial condition plus $-y^2/2t$.}\cite{erik}.
\newline
The extremization procedure implies
\be
x = \hat{y} - t \tanh(\hat{y}) = \hat{y} + u(t,x)t=\hat{y} - \langle m \rangle t,
\ee
where the second equality holds because the Burger equation becomes inviscid in the thermodynamic limit and
which, for $x=0$ (where statistical mechanics is recovered) implies $\hat{y}_0=\langle m \rangle t$ so to obtain the well-known Curie-Weiss self-consistency (properly evaluated by choosing $t=\beta$)
\be
\langle m \rangle = \tanh\left( \beta \langle m \rangle \right).
\ee
The free energy of the Curie-Weiss model is then
\be
\alpha(\beta)= \sup_{\langle m \rangle}\left\{ \ln 2 + \ln\cosh\left( \beta \langle m \rangle \right) - \frac{\beta}{2}\langle m \rangle^2 \right\}.
\ee


While it is well known from classical arguments of statistical mechanics that the Curie-Weiss model undergoes a phase transition from an ergodic (paramagnetic) phase to a ferromagnetic one at $\beta=1$, it is very instructive to tackle this phenomenon still within our framework, where such a phase transition is obtained as a shock wave for the Burger equation (\ref{CWB}).
\newline
In order to see this, it is useful to investigate the mass conservation, whose density is depicted by the variable $\rho$, namely by analyzing the mass of the "fictitious particle" whose motion we study in two different positions, a generic $x$ and the starting point $y$, as
\be
\rho(x) dx = \rho(y) dy.
\ee
By the equation of motion $x = y + u(0,y) t = y - \tanh(y)t$ we get
\be
\frac{dx}{dy}= 1 +\partial_y u(0,y) t= 1- (1-\tanh^2(y))t,
\ee
thus, for the mass density in a generic point $x$, we get 
\be
\rho(x) = \rho(y) \frac{1}{1-(1-\tanh^2(y))t }.
\ee
At $y=0$, $\rho(x)=\rho(0)/(1-t)$ which diverges for $t=1$, thus, as $t=\beta$, exactly where the phase transition happens in statistical mechanics.
\newline
Note further that in the Hamilton-Jacobi equation (\ref{CWHJ}), the potential $V(t,x)$ is vanishing for $N\to\infty$ as $V(t,x)=\partial_{xx}^2 S(t,x)/2N$, and the corresponding Burger equation (\ref{CWB}) becomes inviscid: As a result, by definition of $V(t,x)$, we get self-averaging of the order parameter, namely $\lim_{N \to \infty}(\langle M_N^2 \rangle - \langle M_N \rangle^2)=0$. In the spin-glass counterpart this procedure will develop more complex overlap polynomial identities and, while even in that context we will have $\lim_{N \to \infty}V(t,x)=0$ and $\lim_{N \to \infty}\partial_{xx}^2 S(t,x)/2N=0$, the two results will be in general different, the former representing overlap self-averaging hence restricted only to the RS scenario, the latter representing more general polynomial identities (for simple models as the CW, these two results of course do coincide -as they should- because the model is intrinsically replica symmetric\footnote{For the sake of completeness, note that the Parisi-like representation of the CW model is shown in \cite{BarraCW}.}).

\section{The Replica Symmetric Sherrington-Kirkpatrick model within the Fourier framework.}

Once introduced $N$ Ising spins $\sigma_i = \pm 1$, the Hamiltonian of the SK model is given by
\be
H_N(\sigma;J)= - \frac{1}{\sqrt{N}} \sum_{(i,j)}J_{ij}\s_i\s_j
\ee
where the quenched disorder in the couplings is given by the $N(N-1)/2$ independent and identical distributed random variable $J_{ij}$, whose distribution is $\mathcal{N}[0,1]$.
\newline
We are interested in an explicit expression for the (quenched) free energy $f(\beta)$ (or the mathematical pressure $\alpha(\beta)$) defined as
\be
\alpha(\beta) = -\beta f(\beta) = \lim_{N \to \infty} \alpha_N(\beta) = - \lim_{N \to \infty} \beta f_N(\beta) = \lim_{N \to \infty} \frac1N \mathbb{E} \ln Z_N(\beta),
\ee
where $\mathbb{E}$ averages over the quenched couplings and $Z_N(\beta)=\exp(-\beta H_N(\sigma;J))$ is the partition function.
\newline
Through $Z_N(\beta)$ we define the Boltzmann state $\omega (.)= \sum_{\sigma} . \exp(-\beta H_N(\sigma;J))/(Z_N(\beta))$, the product state $\Omega(.) = \omega(.)\times...\times\omega(.)$ and the averages $\langle . \rangle = \mathbb{E}\Omega(.)$.
\newline
Now, mirroring the previous section, we introduce two fictitious variables $x,t$, which can be though of as space and time coordinates, by which we write the Guerra's interpolating  function as
\be\label{interpolating}
\alpha_N(t,x)= \frac{1}{N} \mb{E} \ln \sum_{\s} \exp\left( \sqrt{\frac{t}{N}}\sum_{i<j}J_{ij}\s_i\s_j + \sqrt{x} \sum_{i}J^1_i\s_i  \right),
\ee
where the $J^1_i$'s are i.i.d. unitary gaussian random variables, and the pressure is recovered whenever evaluating $\alpha_N(t,x)$ at $t=\beta, \ x=0$. Further, as $\alpha$ is not directly connected to an Hamilton-Jacobi equation; we need a linear transformation in the $t,x$ plan to introduce the Guerra's action $S_N(t,x)$ as
\be\label{S}
S_N(t,x)= 2 \alpha_N(t,x) - x - t/2.
\ee
By direct calculation, we can see the following relations holding \cite{sumrule}
\be
\begin{split} \label{d_prim}
\partial_t S_N(t,x) &= - \frac{1}{2} \meanv{q_{12}^2}, \\
\partial_x S_N(t,x) &= - \meanv{q_{12}},
\end{split}
\ee
where we implicitly introduced the overlap, e.g the order parameter of the theory, defined as $q_{12}=N^{-1}\sum_i^N \sigma_i^1 \sigma_i^2$, whose $N$-dependence has been omitted for the sake of simplicity.
By direct construction it is immediate to check that
\be\label{rshj}
\frac{\partial S_N(t,x)}{\partial t} + \frac12\left( \frac{\partial S_N(t,x)}{\partial x}\right)^2 =-\frac12 \left( \langle q^2_{12} \rangle -\langle q_{12} \rangle^2\right)\equiv - V_0(t,x).
\ee
If we add a vanishing (in the thermodynamic limit) potential, containing the second derivative of $S_N(t,x)$
\be
\lim_{N \to \infty}\frac{1}{2N}\frac{d^2S_N(t,x)}{dx^2}\equiv V_1(t,x)=0
\ee
and within the replica symmetric scheme, where $\lim_{N \to \infty}\left( \langle q_{12}^2 \rangle - \langle q_{12} \rangle^2 \right)=0$, $S_N(t,x)$ satisfies
\be
\lim_{N \to \infty}\large( \partial_t S_N(t,x) + \frac12 \left(\partial_x S_N(t,x)\right)^2 - \frac{1}{2N}\partial^2_{x^2} S_N(t,x) \large)=0,
\ee
that we can solve easily with the usual Cole-Hopf transform (see the next section).
\newline
A remark is in order here: As discussed in detail in \cite{BDBG}, and as of course the overlap is not self-averaging in the true solution of the SK model \cite{MPV}, we force $V_0(t,x)$ to be zero in order to get straightforwardly the replica-symmetric solution (which is the goal of the present section), while $V_1(t,x)$ is always zero in the thermodynamic limit (and of course reduces to an elementary identity once read in the RS framework).
\newline
Note that, while it is not strictly necessary to solve this problem where $V_0(t,x)$ and $V_1(t,x)$ are pasted in the same equation as we could split the standard Hamilton-Jacobi equation for the Guerra action from the constraint $\frac{1}{2N} \frac{\partial^2 S_N(t,x)}{\partial x^2}=0$, however, such a ''compact procedure''  allows to obtain the RS free energy solving a Fourier problem (with all its related know-now) for its Cole-Hopf transform.
\newline
To compute explicitly $V_1(t,x)$ it is convenient to introduce the $x$-streaming relative to a generic observable $F$ which depends on $s$ replicas as \cite{sumrule}
\be
\partial_x \meanv{F_s}= N \langle F \left( \sum_{a b}^s q_{a b} - s \sum_{a}^s q_{a s+1} + \frac{s(s+1)}{2}  q_{s+1,s+2} \right)\rangle,
\ee
hence, remembering that, from (\ref{d_prim}), $\partial_x{S(x,t)}= -\langle q_{12} \rangle $, we get 
\be\label{d_sec}
\lim_{N\to \infty}\frac{-1}{N} \partial_{x^2}^2 \, S_N(t,x) = \lim_{N\to \infty}  \large(\meanv{q_{12}^2} - 4\meanv{q_{12}q_{23}} + 3 \meanv{q_{12}q_{34}}\large)=0.
\ee
The latter is a constraint between overlap's polynomials known as AC identities \cite{ac} and in particular it coincides with eq.s $(47,48)$ of \cite{guerra2}, where this kind of identities appeared for the first time, previa elimination of $\langle q_{12} \rangle^2$ to reduce to a single expression.  Hence, as a difference with the CW, $V_0(t,x) \neq V_1(t,x)$ but they both vanish in the $N\to\infty$ limit ($V_0$ only in the RS framework) and they carry different physics inside (as we are going to deepen later, in the broken replica framework).
\newline
As we did for the CW model, we can solve the Burger-like equation for the action
\be
\p_t S_N(t,x) + \frac{1}{2} (\p_x S_N(t,x))^2 - \frac{1}{2N} \p_{x^2}^2 S_N(t,x) =0
\ee
mapping the latter into a Fourier equation via the following Hopf-Cole transform, namely
\be
\Psi_N(t,x)= \exp{(-N S_N(t,x))}
\ee
by which it is straightforward to check that $\Psi_N(t,x)$ obeys:
\be\label{Fourier_RS}
\frac{\partial \Psi(t,x)}{\partial t} - \frac{1}{2N} \frac{\partial^2 \Psi(t,x)}{\partial x^2} =0.
\ee
By calling $\hat{\Psi}_N(t,k)$ the transform of $\Psi_N(t,x)$, in the impulse space we get
\be
\p_t \hat{\Psi}_N(t,k) + \frac{k^2}{2N}\hat{\Psi}_N(t,k)
\ee
hence, using the label $\Psi_0(k)$ to denote the Cauchy condition, we get the solution in the $k$-space as
\be\label{Psi_sol} 
\hat{\Psi}_N(t,k) = \hat{\Psi}_0(k) \exp( - \frac{k^2}{2 N} t ).
\ee
We can obtain the solution in the original space by the Convolution Theorem, hence writing
\be\label{psi}
\Psi(t,x)=  \int d y\, G_t(x-y)  \Psi_0(y) = \sqrt{\frac{N}{2 \pi t}}  \int d y  \,  e^{-N\frac{(x-y)^2}{2t} } \Psi_0(y)
\ee
where $G_t(x-y)$ is the Green propagator, i.e. the solution of (\ref{Fourier_RS}) with the initial condition $G_0(x)= \delta(x)$
and where
\be
\Psi_0(y)= \exp{\Big[-N S_0(y)\Big]}.
\ee
From the definition of the interpolating function (\ref{interpolating}) and from the definition (\ref{S}) one can directly get the expression for $S_0(y)$, namely
$$S_0(y)= 2\ln 2+2 \int d \mu(z) \ln\cosh(\sqrt{y}z)- y,$$
and by direct substitution we get
\be
S_N(t,x) = - \frac{1}{N} \log \Psi (t,x) = - \frac{1}{N} \log\sqrt{\frac{N}{2 \pi t}}  \int d y \, e^{-N\left( \frac{(x-y)^2}{2t} + 2\ln2 + 2 \int d \mu(z) \ln\cosh(\sqrt{y}z)- y\right)}
\ee
that in the thermodynamic limit reads
\be\label{azione_extr} 
\begin{split}
S(t,x) &= \inf_{y}\left\{ \frac{(x-y)^2}{2t} + S_0(y)\right\} \\
&=\inf_{y}\left\{\frac{(x-y)^2}{2t} + 2\ln2 + 2 \int d \mu(z) \ln\cosh(\sqrt{y}z)- y\right\}, \hspace{1cm} \mathrm{for}\, N \to \infty.
\end{split}
\ee
We denote by $\hat{y}(t,x)$ the location where the infimum in (\ref{azione_extr}) is achieved and refer to this function as \emph{inverse Lagrangian function} \cite{erik}. As we are going to show, its inverse $x(t,y)$ is the usual Lagrangian function, or rather, the location at time $t$ of the fictitious  particle initially at $y$. We observe that the previous maximizing condition can also be expressed as:
\be\label{action}
S(t,x)- \frac{x^2}{2t} = -\sup_{y}[ \phi_0(y) + \frac{x y}{t} ]
\ee
where $\phi_0(y)= - y^2/2t -S_0(y)$. Hence, the solution of the Burgers-like equation can be expressed again in terms of a Legendre transform of $\phi_0(y)$. From the extremization condition  we get
\be\label{eq_moto}
x= \hat{y} - t \int d \mu(z) \tanh^2(\sqrt{\hat{y}}z) = \hat{y}+ t \, u(t,x),
\ee
with $\hat{y}$ maximizer. The last equality is allowed because in the thermodynamic limit the Burger equation becomes inviscid, hence trajectories represent Galilean motion with a velocity $u(t,x)$ given explicitly by the previous expression.
\newline
Under the \emph{replica symmetric assumption}, from equation (\ref{d_prim}) and the definition of $u(t,x)=\partial_x S(t,x)$, in the thermodynamic limit we can finally recover the self-consistent equation for the overlap
\be
\begin{split}
u(t,x)= \partial_x S(t,x) = -\meanv{q_{12}}(t,x)=-\bar{q}(t,x)=- \int d \mu(z) \tanh^2(\sqrt{\hat{y}(t,x)}z).
\end{split}
\ee
To recover statistical mechanics we need evaluating observables at $x=0$ (e.g. in equation (\ref{eq_moto})) and the value of $\hat{y}_0=t\bar{q}$  that maximizes  the expression (\ref{action}) is then
\be
\hat{y}_0= t \, \bar{q}= t \int d \mu(z) \tanh^2(\sqrt{t\bar{q}}z) .
\ee
By considering equation (\ref{azione_extr}), with  $\hat{y}$ maximizer, we have
\be
S(x,t) = \frac{(x-\hat{y})^2}{2t} + 2\ln2 + 2 \int d \mu(z) \ln\cosh(\sqrt{\hat{y}}z)- \hat{y},
\ee
and when evaluating everything at $x=0$ and $t=\beta^2$ (thus we use the relation $\hat{y}_0 =\bar{q} \,\beta^2$) we finally get
\be
\alpha(\beta)= \ln2 +  \int d \mu(z) \ln\cosh(\sqrt{\bar{q}}\,\beta z)+ \frac{\beta^2}{4}(1-\bar{q})^2
\ee
which is the expression of the pressure for the SK model in the RS approximation.


As in the Curie-Weiss model, we want to investigate the shock waves in the Burgers equation and figure out some correspondences with the phase transition in statistical mechanics (which is known to happen at $\beta=1$). To achieve this result, mirroring the corresponding section for the ferromagnetic case, we consider the conservation of the mass for a particle which starts the motion in position $y$ and arrives at position $x$ after a time $t$:
\be
\rho(x) dx = \rho(y) dy.
\ee
Reminding the equation of motion $x = y + u(0,y) t $, from mass conservation we get
\be\label{eq_density}
\rho(x) = \rho(y) \Big (\frac{dx}{dy}\Big)^{-1}= \frac{\rho(y)}{1 +t\, \partial_y u(0,y)}.
\ee
From equation $(\ref{eq_moto})$ we get
\be
\frac{dx}{dy}= 1 +t \ \partial_y u(0,y)= 1- t \, \frac{1}{\sqrt{y}}\int d \mu(z) z \frac{\tanh(\sqrt{y} \ z)}{\cosh^2{(\sqrt{y}z)}}\bigg.
\ee
If we want to find a shock we can evaluate $\min_{y} \Big[- \, \frac{1}{\sqrt{y}}\int d \mu(z) z \frac{\tanh(\sqrt{y} \ z )}{\cosh^2{(\sqrt{y}z)}}  \Big]$, which can be achieved simply observing parity and  monotony in $y \in [0,\infty)$ hence obtaining $\bar{y}=0$. Then if we expand the $\tanh$ for small $y$ we obtain
\be
\min_{y} \Big[- \, 2 \int_{0}^{\infty} d \mu(z) \frac{z^2}{\cosh^2{(\sqrt{y}z)}} + \frac{2}{3} y \ \int_{0}^{\infty} d \mu(z) \frac{z^4}{\cosh^2{(\sqrt{y}z)}}\Big] = -1,
\ee
where the final result is obtained computing the argument inside the brackets for $y=\bar{y}=0$. We can then conclude that $\partial_y u(0,y)|_{\bar{y}}= -1$. If we now substitute this last result in eq (\ref{eq_density}) we get
\be%
\rho(x) = \frac{\rho(\bar{yn})}{1 +\partial_y u(0,y)|_{\bar{y}} t} =  \frac{\rho(0)}{1 - t}
\ee
which diverges at the shock time $t=\beta^2=1$, i.e. the critical noise level at which the phase transition occurs in statistical mechanics.

\section{Beyond the replica symmetric scenario: Multiple diffusion and broken replica constraints.}

In this section we want to go beyond the replica-symmetric scenario and investigate features of the broken replica phase by merging the PDE approach, in particular the Fourier technique, with the broken replica interpolation scheme developed by Guerra in \cite{broken}. The result will be a mapping between the SK free energy with some steps of RSB (that we are going to analyze in full detail for the $1-RSB$ test-case for the sake of simplicity, mirroring the work developed in \cite{BDBG} for the Hamilton-Jacobi procedure), and a multi-dimensional diffusion equation, with as spatial dimensions as the number of RSB steps plus one, hence two spatial dimensions in order to tackle the $1-RSB$ solution.
\newline
To accomplish this task let us introduce the following interpolating partition function
\be\label{Z1RSB}
Z_N(t,x_1,x_2) = \sum_{\sigma}\exp\left(\sqrt{t}H_N(\sigma;J) + \sqrt{x_1}\sum_i^N J_i^1 \sigma_i + \sqrt{x_2-x_1} \sum_i^N J_i^2 \sigma_i \right),
\ee
where $J_i^1, \ J_i^2$ are all i.i.d. Gaussian variables, sharing the same distribution $\mathcal{N}[0,1]$ as the original $J_{ij}$ but independent from them, whose averages will be denoted with the pertinent subscript for the sake of clearness, hence $\mathbb{E}_2, \ \mathbb{E}_1$. Moreover we have to think at $m\in[0,1]$ as the expected parameter of the 1-RSB overlap's distribution, i.e. we assume its shape to be the weighted sum of two delta functions,  $P(q)=m\delta(q-\bar{q}_1)+(1-m)\delta(q-\bar{q}_2)$.
\newline
Let us introduce recursively the ''partially averaged" partition functions as
$$
Z_2 \equiv Z_N, \ \ Z_1^m = \mathbb{E}_2(Z_2^m),
$$
by which we can express the 1-RSB SK free-energy (or pressure strictly speaking) in the space $(t,x_1,x_2)$ as
\be
\alpha(t,x_1,x_2)=\lim_{N \to \infty}\alpha_N(t,x_1,x_2)= \lim_{N \to \infty}\frac{1}{N}\mathbb{E}\mathbb{E}_1\log Z_1 (t,x_1,x_2) = \lim_{N \to \infty} \frac{1}{N m} \mathbb{E}\mathbb{E}_1\log \mathbb{E}_2 Z_N^m(t,x_1,x_2).
\ee
Trough the interpolating structure defined by the extended partition function (\ref{Z1RSB}) and the partial averages $\mathbb{E}_j$, we can introduce also the following weights $f$ and the corresponding extended states $\omega$ as
\begin{eqnarray}
f_1=1 , \ \ &&f_2=\frac{Z_2^m}{\mathbb{E}_2 Z_2^m},  \\
\omega_1(.) = \mathbb{E}_2[f_2\omega_2(.)], \ \ &&\omega_2(.) = \omega_N(.),
\end{eqnarray}
while the standard product state remains  $\Omega_a(.) = \omega_a(.) \times ... \times \omega_a(.)$  and finally the quenched averages read
\be
\langle . \rangle_a = \mathbb{E}\left( f_1 ... f_a \Omega_a(.) \right),
\ee
hence $\langle . \rangle_1 = \mathbb{E}[\Omega_1(.)]$, and $\langle . \rangle_2 = \mathbb{E}[f_2 \Omega_2(.)]$.
\newline
If we now define
\be\label{defs}
S_N(t,x_1,x_2)= 2 \alpha_N(t,x_1,x_2) - x_2 - t/2,
\ee
we can see that the following relations hold:
\begin{eqnarray}\nonumber
\partial_t S_N(t,x_1,x_2) &=& - (m/2) \langle q_{12}^2 \rangle_1 - ((1-m)/2) \langle q_{12}^2 \rangle_2, \\ \nonumber
\partial_{x_2} S_N(t,x_1,x_2) &=& -(1-m)\langle q_{12} \rangle_2, \\ \nonumber
\partial_{x_1} S_N(t,x_1,x_2) &=& -m \langle q_{12} \rangle_1,
\end{eqnarray}
by which we can see that  $S_N(t,x_1,x_2)$ satisfies
\begin{eqnarray}\label{rsbhj}
\partial_t S_N + \frac{1}{2m}\left(\partial_{x_1}S_N\right)^2 + \frac{1}{2(1-m)}\left(\partial_{x_2}S_N \right)^2
= -\frac 1 2 \left[m\left(\left\langle q^2_{12}\right\rangle_1-\left\langle q_{12}\right\rangle^2_1\right)+(1-m)\left(\left\langle q^2_{12}\right\rangle_2-\left\langle q_{12}\right\rangle_2^2\right)\right].
\end{eqnarray}
In the 1-RSB approximation we can neglect the r.h.s. of ($\ref{rsbhj}$), moreover, if we add a vanishing term containing the second derivatives of $S_N(t,x_1,x_2)$ we can write down a Burger equation for $S(t,x_1,x_2)$ as
\begin{eqnarray}\label{rsbb}
&&\partial_t S_N + \frac{1}{2m}\left(\partial_{x_1}S_N\right)^2 + \frac{1}{2(1-m)}\left(\partial_{x_2}S_N \right)^2= \frac 1 {2Nm}\partial^2_{x_1^2}S_N +\frac 1 {2N(1-m)}\partial^2_{x_2^2}S_N
\end{eqnarray}
that we can solve  by mapping it into a Fourier equation via the usual Cole-Hopf transform
\be
\Psi_N(t,x_1,x_2)= \exp\left( - N S_N(t,x_1,x_2) \right),
\ee
by which it is straightforward to check that $\Psi_N(t,x_1,x_2)$ obeys
\be\label{fourier1RSB}
\lim_{N \to \infty} \large( \partial_t \Psi_N(t,x_1,x_2) -  \frac{1}{2 N m} \partial_{x_1^2}^2 \Psi_N(t,x_1,x_2) - \frac{1}{2 N (1-m)} \partial^2_{x_2^2} \Psi_N(t,x_1,x_2) \large) = 0,
\ee
i.e. a diffusion equation with two different diffusion coefficients on the two spatial axes $x_1, \ x_2$, namely $D_1 = (1/2 N m)$ and $D_2 = 1/(2 N (1- m))$.
\newline
Now we want to solve the heat-equation (\ref{fourier1RSB}) in the Fourier space, where, calling $\hat{\Psi}_N(t,k_1,k_2)$ the transform of $\Psi_N(t,x_1,x_2)$ we can write
\be
\partial_t \hat{\Psi}_N(t,k_1,k_2) + \frac{k_1^2}{2 m N} \hat{\Psi}_N(t,k_1,k_2) + \frac{k_2^2}{2 (1-m) N} \hat{\Psi}_N(t,k_1,k_2) = 0,
\ee
hence, using the label $\Psi_0(k_1,k_2)$ to denote the Chauchy condition, we get
\be
\hat{\Psi}_N(t,k_1,k_2) = \hat{\Psi}_0(k_1,k_2) \exp\left( - \left(  \frac{k_1^2}{2 m N}  + \frac{k_2^2}{2 (1-m) N}  \right) t  \right).
\ee
Finally we can obtain the solution in the original space by the Convolution Theorem, hence writing
\begin{eqnarray}
\Psi_N(t,x_1,x_2) &=& \int \frac{d k_1}{\sqrt{2\pi}} \int \frac{d k_2}{\sqrt{2\pi}} \hat{\Psi}_0(t,k_1,k_2) \exp\left( - \left(  \frac{k_1^2}{2 m N}  + \frac{k_2^2}{2 (1-m) N}  \right) t  \right) \exp(i k_1 x_1 + i k_2 x_2)\\
&=&  \int d y_1 \int d y_2 \Psi_0(t,y_1,y_2) G_t(x_1-y_1,x_2-y_2),
\end{eqnarray}
where $G_t(x_1-y_1,x_2-y_2)$ is the Green propagator, i.e. the solution of (\ref{fourier1RSB}) with the initial condition $G_0(x_1,x_2)=\delta(x_1)\delta(x_2)$.
\be
G_t(x_1,x_2) = \int \frac {d k_1}{2 \pi} \int \frac {d k_2}{2 \pi} e^{-\left( \frac{k_1^2}{2 m N}+\frac{k_2^2}{2(1-m)N} \right) t } e^{(i k_1 x_1 + i k_2 x_2)}= \frac{N}{2\pi t} e^{-N\left( \frac{m}{2t}x_1^2 + \frac{(1-m)}{2t}x_2^2 \right)},
\ee
by which the solution of eq. (\ref{fourier1RSB}) is
\be
S_N(t,x_1,x_2)= -\frac1N \log \Psi(t,x_1,x_2)= -\frac{1}{N}\log \int dy_1 \int dy_2 e^{-N\left( S_0(y_1,y_2)+ \frac{m}{2t}(x_1-y_1)^2 + \frac{(1-m)}{2t}(x_2-y_2)^2 \right)}.
\ee
In the thermodynamic limit we can use the saddle point method to get
\be\label{srsb}
S(t,x_1,x_2)=\inf_{y_1,y_2}\left\{ S_0(y_1,y_2) + \frac{m}{2t}(x_1-y_1)^2 + \frac{(1-m)}{2t}(x_2-y_2)^2 \right\},
\ee
and finally, keeping in mind the definition $(\ref{defs})$
\be\label{a1RSB}
\alpha(t,x_1,x_2)=\inf_{y_1,y_2}\left\{ \frac12 S_0(y_1,y_2)+ \frac{m}{4t}(x_1-y_1)^2 + \frac{(1-m)}{4t}(x_2-y_2)^2 + \frac{x_2}{2} + \frac{t}{4}\right\}.
\ee
At $t=0$ we have
\be
S_0(y_1,y_2) = 2\alpha(0,y_1,y_2)-y_2=2\log 2+\frac{2}{m}\int d\mu(z_1)\log \int d\mu(z_2)\cosh^m \left( \sqrt{y_1}z_1 + \sqrt{y_2-y_1}z_2 \right) - y_2.
\ee
If the minimum of the equation $(\ref{srsb})$ is achieved in $(y_1,y_2)=(\hat{y}_1,\hat{y}_2)$, we can write
\begin{eqnarray}
\frac{d S}{d x_1}(t,x_1,x_2) &=&- m \langle q_{12} \rangle_1= m \frac{x_1 - \hat{y}_1(t,x_1,x_2)}{t}, \\
\frac{d S}{d x_2}(t,x_1,x_2) &=&- (1- m) \langle q_{12} \rangle_2= (1-m) \frac{x_2 - \hat{y}_2(t,x_1,x_2)}{t},
\end{eqnarray}
by which we get the physical meaning of the variables $\hat{y}_1, \hat{y}_2$ as
\begin{eqnarray}
\hat{y}_1(t,x_1,x_2) &=& x_1 + \langle q_{12}(x_1,x_2,t)\rangle_1 t, \\
\hat{y}_2(t,x_1,x_2) &=& x_2 + \langle q_{12}(x_1,x_2,t)\rangle_2 t.
\end{eqnarray}
This means that, at $t=\beta^2$ and $x_1=x_2=0$, where we come back to the original SK model, $(\hat{y}_1,\hat{y}_2)=\beta^2 (\langle q_{12}\rangle_1,\langle q_{12}\rangle_2)=\beta^2(\bar{q}_1,\bar{q}_2)$.


Now we want to extremize the expression inside the brackets of  (\ref{a1RSB}). Taking the derivatives w.r.t. $y_1$ and $y_2$, by straightforward calculation we get
\begin{eqnarray}
\frac{x_1 - \hat{y}_1(t,x_1,x_2)}{t}&=&  \int d \mu(z_1) \left[D^{-1}(z_1,\hat{y})\int d \mu (z_2) \cosh^m(\Theta(\hat{y},z_1,z_2))\tanh(\Theta (\hat{y},z_1,z_2))\right]^2\nonumber\\
\frac{x_2 - \hat{y}_2(t,x_1,x_2)}{t} &=& \int d \mu(z_1) D^{-1}(z_1,\hat{y})\int d \mu (z_2) \cosh^m(\Theta(\hat{y},z_1,z_2))\tanh^2(\Theta (\hat{y},z_1,z_2)),
\end{eqnarray}
where we  defined
\begin{eqnarray}
D(z_1)&=& \int d \mu(z_2) \cosh^m(\sqrt{\hat{y}_1}z_1 + \sqrt{\hat{y}_2-\hat{y}_1}z_2)\nonumber\\
\Theta(\hat{y},z_1,z_2)&=&\sqrt{\hat{y}_1}z_1 + \sqrt{\hat{y}_2-\hat{y}_1}z_2,
\end{eqnarray}
and by which we can reconstruct the function $\alpha(t,x_1,x_2)$ as
\begin{eqnarray}
\alpha(t,x_1,x_2)&=& \ln 2 + \frac{m}{4t}(x_1-\hat{y}_1(t,x_1,x_2))^2 + \frac{1-m}{4t}(x_2-\hat{y}_2)^2 + \frac{x_2-\hat{y}_2}{2}+\frac{t}{4}\nonumber \\
&+& \frac{1}{m}\int d \mu(z_1) \ln \int d \mu (z_2) \cosh^m\left( \sqrt{\hat{y}_1}z_1 + \sqrt{\hat{y}_2 - \hat{y_2}}z_2\right).
\end{eqnarray}
At $t=\beta^2$, $x_1=x_2=0$ and identifying $(\hat{y}_1,\hat{y}_2) = \beta^2(\bar{q}_1,\bar{q}_2)$ we find the following well known \cite{broken}\cite{BDBG}  self-consistent equations
\begin{eqnarray}\label{selfq}
\bar{q}_1&=&\int d \mu (z_1)\left[ D^{-1}(z_1, \bar{q})\int d \mu(z_2) \cosh^m(\beta \sqrt{\bar{q}_1} z_1 + \beta \sqrt{\bar{q}_2-\bar{q}_1} z_2) \tanh(\beta \sqrt{\bar{q}_1}z_1 + \beta \sqrt{\bar{q}_2 - \bar{q}_1}z_2)\right]^2\nonumber\\
\bar{q}_2&=&\int d \mu (z_1) D^{-1}(z_1, \bar{q})\int d \mu(z_2) \cosh^m(\beta \sqrt{\bar{q}_1} z_1 + \beta \sqrt{\bar{q}_2-\bar{q}_1} z_2) \tanh^2(\beta \sqrt{\bar{q}_1}z_1 + \beta \sqrt{\bar{q}_2 - \bar{q}_1}z_2)
\end{eqnarray}
and the $1-RSB$ free energy of the SK model as
\be
\alpha(\beta) = \ln 2 + \frac{\beta^2}{4}\left( m \bar{q}_1^2 + (1-m)\bar{q}_2^2 - 2 \bar{q}_2+1\right) + \frac{1}{m}\int d \mu(z_1) \ln \int d \mu(z_2) \cosh^m \left( \beta \left( \sqrt{\bar{q}_1}z_1 + \sqrt{\bar{q}_2-\bar{q}_1}z_2 \right)  \right).
\ee

Earlier, in writing the equation $(\ref{rsbb})$, we added a vanishing (in the thermodynamic limit) term concerning the second derivatives of $S(t,x_1,x_2)$:
\begin{eqnarray}\label{vanterm}
-\frac{1}{Nm}\partial^2_{x_1^2}S&=&\frac 1 N \partial_{x_1} \langle q_{12} \rangle_1 \stackrel{N\to\infty}{\rightarrow} 0, \\ \label{vanterm2}
\frac{1}{N(m-1)}\partial^2_{x_2^2}S&=&\frac 1 N \partial_{x_2} \langle q_{12} \rangle_2 \stackrel{N\to\infty}{\rightarrow} 0.
\end{eqnarray}
These terms are vanishing because $S(t,x_1,x_2)$ is a continuous function in its parameters and its thermodynamic limit exists due to the Guerra-Toninelli argument \cite{limterm} or, from a more practical point of view, as we have found the correct 1-RSB free energy, they necessarily have to be  vanishing in the thermodynamic limit. Analyzing the equations ($\ref{vanterm}, \ref{vanterm2}$) allows us to find some overlap constraints that generalize the AC like identities found inside the RS scheme (equation ($\ref{d_sec}$)).
Let us start exploiting the meaning of the first equation. Let us stress that $\Omega_2(.)=\omega_2(.)\times...\times\omega_2(.)=\Omega(.)$, while $\Omega_1(.)=\omega_1(.)\times...\times\omega_1(.)$, furthermore, for the sake of clearness, let us remember that $\langle . \rangle_2 = \mathbb{E}[f_2 \Omega_2(.)]$ and $\langle . \rangle_1 = \mathbb{E}[\Omega_1(.)]$. We need further the introduction of the state
$$
\tilde{\Omega}_1 = \mathbb{E}_{2}[f_{2}\Omega(.)],
$$
that is not a replicated state because couples different replicas with the $ \mathbb{E}_{2}$ average. Finally we have to define composite states: without using an unnecessarily general notation, we will define for example
\be
\left\langle .\right\rangle_{\tilde{\Omega}_1\times\Omega_1}=\mathbb{E}[(\tilde{\Omega}_1\times\Omega_1)(.)]
\ee
and similarly for any other possible combination of replicated states.
It is convenient to introduce the following preliminary results:
\begin{eqnarray}
\partial_{x_2} f_2 &=& \partial_{x_2} \frac{Z^m}{\mathbb{E}Z^m} =\frac{1}{2\sqrt{x_2-x_1}}\sum_k m f_2 \left( J^2_k\omega(\sigma_k) - \mathbb{E}_2 [J^2_k f_2 \omega(\sigma_k)]  \right), \\
\partial_{J_i^2}f_2 &=& m \frac{Z^m}{\mathbb{E}Z^m}Z^{-1} \partial_{J_i^2}Z = m f_2 \omega(\sqrt{x_2-x_1}\sigma_i),\\
\partial_{J_i^2}\omega(F) &=& \sqrt{x_2-x_1}\left( \omega(F \sigma_j)-\omega(F) \omega(\sigma_j) \right),\\
\partial_{x_2}\omega(F) &=& \frac{1}{2\sqrt{x_2-x_1}}\sum_k J^2_k \left( \omega(F \sigma_k)- \omega(F) \omega(\sigma_k) \right).
\end{eqnarray}
With these premises, the evaluation of the streaming of $\langle q_{12} \rangle_2$ over $x_2$ can be written as
\be
\partial_{x_2}\langle q_{12} \rangle_{2} = \mathbb{E}\left( \partial_{x_2}f_2 \Omega(q_{12}) \right) + \mathbb{E}\left( f_2 \partial_{x_2}\Omega(q_{12}) \right) = \mathfrak{A}+\mathfrak{B},
\ee
where
\begin{eqnarray}
\mathfrak{A} &=& N \left( m \langle q_{12}q_{23} \rangle_2 + \frac{m(m-3)}{2}\langle q_{12}q_{34}\rangle_2 + \frac 1 2 m(1-m)2\langle q_{12}q_{34}\rangle_{\tilde{\Omega}_1\times\tilde{\Omega}_1} \right),\\
\mathfrak{B} &=& N \large( (m-4)\langle q_{12}q_{23}\rangle_2 + \langle q_{12}^2 \rangle_2 + (3-m)\langle q_{12}q_{34} \rangle_2  \large),
\end{eqnarray}
hence, finally the first constraint reads off as
\be
\lim_{N \to \infty}\frac 1 N \partial_{x_2}\langle q_{12}\rangle_2 = 0 = \langle q_{12}^2 \rangle_2 + 2(m-2)\langle q_{12}q_{23}\rangle_2 + \frac 1 2(m-3)(m-2)\langle q_{12}q_{34} \rangle_2 + \frac 1 2 m(1-m)2\langle q_{12}q_{34}\rangle_{\tilde{\Omega}_1\times\tilde{\Omega}_1}.
\ee
In the same way we get the other constraint that follows from equation $( \ref{vanterm2})$ by which, overall, we get
\begin{eqnarray}\label{new1}
0 &=& \langle q_{12}^2 \rangle_2 + 2(m-2)\langle q_{12}q_{23} \rangle_2 + \frac 1 2(m-2)(m-3) \langle q_{12}q_{34} \rangle_2 + \frac 1 2 m(1-m)\left\langle q_{12}q_{34}\right\rangle_{\tilde{\Omega}_1\times\tilde{\Omega}_1},\\ \label{new2}
0 &=& \langle q_{12}^2 \rangle_1 - 4 m \langle q_{12}q_{23}\rangle_1 + 3m^2 \langle q_{12}q_{34}\rangle_1+2(m-1)\left\langle q_{12}q_{13}\right\rangle_{\omega_1\times\tilde{\Omega}_1}\nonumber\\
&+&4m(1-m)\left\langle q_{13}q_{24}\right\rangle_{\Omega_1\times\tilde{\Omega}_1}+(1-m)^2\left\langle q_{13}q_{24}\right\rangle_{\tilde{\Omega}_1\times\tilde{\Omega}_1}+2m(m-1)\left\langle q_{12}q_{34}\right\rangle_{\Omega_1\times\tilde{\Omega}_1}
\end{eqnarray}
Note that, as expected, when $m=0$ the first constraint becomes as the AC like identity ($\ref{d_sec}$), derived using the RS interpolating scheme, but concerning the state $\left\langle .\right\rangle_1$, i.e. averaging only inside the first ensemble of valleys. On the contrary, when $m=1$, the second constraint becomes the AC like identity ($\ref{d_sec}$) but with the state $\left\langle .\right\rangle_2$, i.e. averaging just inside the second ensemble of valleys.  When $m\in(0,1)$ other terms appear to take into account the relation between the two valleys, hence constraining, for the first time, this kind of overlap correlations.

\section{Conclusion and outlooks}

In this paper we highlighted how standard PDE techniques, well consolidated in different fields of mathematical research (e.g. classical mechanics and hydrodynamics), may work as well once tested on the disordered statistical mechanics machinery. We extended previous results \cite{sumrule}\cite{BarraCW}\cite{BDBG}\cite{BG1} on this line, showing how to solve trough Fourier theory for the free energy and via a Cole-Hopf transform by using Burger theory for the order parameter. Remarkably, as the latter admits shock waves, we found that -within this parallel- those happen exactly where phase transitions develop in standard statistical mechanics.
\newline
Beyond showing the main ideas on the Curie-Weiss model for the sake of clearness, we tested our approach to the paradigmatic Sherrington-Kirkpatrick model, which has been solved both within a replica symmetric scenario and within the Guerra broken replica framework, at the first level of replica symmetry breaking for the sake of simplicity, finding full agreement with the classical theory.
\newline
We remark however that, while general $K$-steps of RSB make the calculations only longer, but work straightforwardly, the $K\to\infty$ limit requires still some effort on which we will deserve future investigations.
\newline
A main point to highlight is that our procedure is a novel scheme to produce (polynomial) overlap constraints, which have been recently found to play a crucial role \cite{panchenko1,panchenko2} in the ultrametric organization of the states predicted by Parisi theory \cite{MPV} and that we obtain as a constraint on the theory. To get an entirely class of new identities, we used a combination of results: as those which could be derived within the standard Gibbs measure -in the fully connected topology underlying the SK model- have been already obtained (and we found them as well in the simpler RS-framework) we worked out this kind of constraints within the Guerra broken replica weights. This procedure resulted in correlations that link overlap fluctuations between different measures, which roughly resembles constraints among valleys of different depth (corresponding to different steps of RSB from an heuristic perspective). We remark that these constraints have been obtained very naturally, from continuity arguments, as the second (spatial) derivative of the action, once divided by the volume, must go to zero in the thermodynamic limit: Roughly speaking, this approach resembles self-averaging in older investigations \cite{gg},  and, forcing  the broken-replica scheme to collapse on the simple Gibbs measure (hence imposing $m=1$),  results reduce to those well known polynomial identities (already found in \cite{ac,guerra2,barra1}). Interestingly in this way, we obtained an entirely new class of identities that constraint overlap fluctuations within {\em different} levels of RSB: The implication in possible closure of the SK-repertoire, hence in supporting Panchenko results will be subject of future investigation as well.

\section*{Acknowledgements}

The authors are indebted with Erik Aurell for several illuminating discussions on the shock wave and the inverse Lagrangian.
\newline 
Francesco Guerra is warmly acknowledged as usual for his priceless constant guide.
\newline
AB and DT acknowledge the FIRB grant $RBFR08EKEV$ and Sapienza Universita' di Roma for partial financial support and GDF acknowledge a grant by Netadis Project.

\end{document}